# METHODOLOGICAL PRINCIPLES OF MODERN THERMODYNAMICS

Etkin V.A.

The article describes basic principles of the theory which unites
thermodynamics of reversible and irreversible processes also extends them
methods on processes of transfer and transformation of any forms of energy

**Introduction.** There are periods in the development of any natural-science theory when new ideas and experimental facts can not be crammed into "Procrustean bed" of its obsolete notional and conceptual system. Then the theory itself – its presuppositions, logical structure and body of mathematics – becomes the object of investigation. Thermodynamics went through such periods more than once [1]. So was yet in the mid-XIX century when under the pressure of new experimental facts the concept of heat as an indestructible fluid collapsed and "entrained" (as seemed then) the S. Carnot's theory of heat engines based on it [2]. A few decades later the threatening clouds piled up over the R. Clausius' mechanical theory of heat because of the "heat death of the Universe" – a conclusion deemed then as inevitable [3].

In the late XIX century great difficulties arose from attempts to conduct a thermodynamic analysis of composition variation in heterogeneous systems (at diffusion, chemical reactions, phase transitions, etc). J. Gibbs [4] overcame the majority of those difficulties by representing closed system as a set of open subsystems (phases and components), which allowed him to reduce the internal processes of system composition variation to the external mass transfer processes. However, some of those difficulties have remained as yet and are showing, in particular, in the unsuccessful attempts to thermodynamically resolve the "Gibbs' paradox" – a conclusion of stepwise entropy rise when mixing non-interacting gases and independence of these steps on the nature the gases feature and the degree they differ in [5…7].

During the XIX century thermodynamics also more than once encountered paradoxical situations that arose around it with the human experience outstepped. One of such situations arose with thermodynamics applied to the relativistic heat engines (contain fast moving heat wells) and showed in the statement that those could reach efficiency higher than in the Carnot's reversible engine within the same temperature range [8…11], as well as in the recognized ambiguity of relativistic transformations for a number of thermodynamic values [7]. A little bit later a situation, not any less paradoxical, arose as connected with attempts to thermodynamically describe the systems of nuclear magnets (spin systems) with inverted population of energy levels. The negative absolute temperature concept introduced for such states led investigators to a conclusion of possibility for heat to completely convert into work in such systems and, on the contrary, impossibility for work to completely convert into heat, i.e. to the "inversion" of the principle fundamental for thermodynamics – excluded perpetual motion of the second kind [7,11…13].

That fate became common for also the theory of irreversible processes (TIP) created by extrapolating classic thermodynamics to non-equilibrium systems with irreversible (non-static) processes running therein. Problems arose primarily from the necessity to introduce into thermodynamics the transfer concepts inherently extraneous for it, from the incorrectness to apply the equations of equilibrium thermodynamics to irreversible processes in view of their inevitable change to inequalities, from the inapplicability of the classic notions of entropy and absolute temperature to thermally heterogeneous media, etc, which demanded to introduce a number of complimentary hypotheses and to attract from outside balance equations for mass, charge, momentum, energy and entropy with time involved as a physical parameter. Even heavier obstructions arise with attempts to generalize TIP to non-linear systems and states far away from equilibrium where the Onsager-Casimir reciprocal relations appear to be violated [14,15] and the law of entropy minimal production becomes invalid [16,17]. Attempts to overcome these difficulties without whatever correction on the conceptual fundamentals and body of mathematics of classic thermodynamics failed.

A remedy can be found in building modern thermodynamics on its own more general notional and conceptual foundation with maximal care for the classic thermodynamic heritage. Conceptual bases of such theory are discussed more low.

### 1.1. System Approach to Objects of Investigation.

The intention to keep the advantages of the thermodynamic method when generalizing it to non-thermal forms of energy and non-static processes impelled us to base modern thermodynamics as consistently phenomenological and deductive theory. The deductive method (from the general to the particular) is much closer to *the system* approach accepted nowadays for the standard which assumes the account of all correlations. Its basic difference is part studying through whole (rather the reverse).

Hardly it is necessary to prove, as far from it other fundamental disciplines adhering to a return (inductive) method of research. This method assumes possibility of studying whole through its elementary parts by summation of their extensive properties. With that end in view investigated system is split up on huge (in case of a continuum – infinite) number of elementary volumes, material points, elementary particles, etc., assumed internally equilibrium (homogeneous). It deprives investigated objects of the basic property of material bodies - their *spread*, and any lengthy object – *the structure* caused by spatial heterogeneity of object of research.

However, the extensive properties of heterogeneous systems are far from being always additive ones, i.e. the sum of properties of constituent elements. First of all, non-additive is the property of a heterogeneous system to do useful work as none of its local parts possesses it. It was S. Carnot who awoke to that statement in application to heat engines (1824) and put it into historically the first wording of the second law of thermodynamics. According to it, only thermally heterogeneous media possess a "vis viva (living force)", i.e. are able to do useful work. In itself the notion of *perpetual motion of the second kind* as a system with no heat well and heat sink in its structure evidences the importance of considering such media as a single whole (but not as a set of thermally homogeneous elements). This is just the reason, why, at study of heat engines, the so-called "extended" systems have to be considered, which include, along with heat wells (sources), also heat sinks (receivers) (the environment).

Another non-additive property of heterogeneous media is the internal relaxation processes progressing and resulting, in absence of external constraint, in the equalization of densities, concentrations, electric charges, etc. among various parts of such a system. These processes are, however, absent in any element of the continuum considered as a locally equilibrium part of the system.

One more non-additive property is the impulse of system which can be absent in system as whole, being distinct from zero for any its part. In the same way are non-additive the forces of a gravitational or electrostatic attraction proportional to product of co-operating mass or charges.

More non-additive properties are the *self-organization* ability of a number of systems, which is absent in any of their homogeneous part [18…20], as well as the *synergism* (collective action) phenomena appearing at only a definite hierarchic level of the system organization. The said refers in general to any structured systems, which specific features are determined by the inter-location and inter-orientation of the functionally detached elements and disappear with decomposition of the object of investigation into these elements [21]. Many of such elements (e.g., macromolecules and cells) being detached remain, however, spatially heterogeneous (locally non-equilibrium) despite their microscopic size (constituting *microcosmos* of a kind). This demands them being approached in the same way as the "extended" macro-systems.

Non-additivity of properties may also manifest itself in the so-called "scale factor" consisting in loosing some key characteristics of an object at its modeling and being presently a close-attention-focus in the theory of fractals. Therefore to export test results from a test model to a full-scale installation, e.g. a chemical reactor, is far from being always feasible.

Lastly, the internal energy of a system in itself is non-additive when fragmentizing the last into ever-diminishing parts due to surface or long-range forces presenting and exerting on these parts. Potential energy of these forces belongs to the entire set of interacting (inter-moving) bodies or body parts and may not be referred to the internal energy (self-energy) of any of them taken separately. This is known to have restricted the applicability of thermodynamics both "from below" and "from above" enforcing consideration of the thermodynamic systems just as "simple" systems where these forces may be neglected [22]. This restriction refers, strictly speaking, to also the external energy of a system, which may be attributed to one of the field-forming bodies just in particular cases (within the applicability of such a notion as "test" charge or "test" mass). In particular, momentum variation in some part of a heterogeneous system is inevitably followed by the same of equal magnitude but opposite sign in its other parts. Therefore to restrict oneself to considering the properties of a particular part of the system is far not always applicable.

In sum, one may conclude that investigating elements of continuum may far not always enable sound judgment of its properties in whole. It is the system approach that allows for all interrelations between the whole and the parts. Such an approach dictates a necessity of considering spatially heterogeneous media as *a non-equilibrium single whole* in all cases when a doubt arises in validity of fragmentizing the object of investigation into some parts. Thus modern thermodynamics considers such a set of the interacting bodies as a primary object of investigation, which may be construed as an isolated (self-contained) system. Scale of the system depends on its heterogeneity rate and covers a wide range of material objects – from nano- to mega-world. Herein, what is considered in classic thermodynamics as the "extended" system (comprising both heat sources and heat receivers) becomes just its part (subsystem) in the modern thermodynamics.

The system approach allows discovering a special class of processes existing in heterogeneous systems and featuring opposite directivity. To demonstrate this, let us compare the density $\rho_i = d\Theta_i/dV$ of any extensive parameter $\Theta_i$ (mass $M$, entropy $S$, charge $\Theta_e$, number of moles $N_k$ in the $k$th substance, etc.) in some system part with its mean integral value $\bar{\rho}_i = V^{-1}\int\rho_i dV = \Theta_i/V$. Then it is easy to mark out areas in the system with volumes of $V'$ and $V''$ where the densities $\rho_i'(\mathbf{r},t)$ and $\rho_i''(\mathbf{r},t)$ are higher and lower than $\bar{\rho}_i$, respectively. Indeed, dividing such a system into areas with the volumes $V'$ and $V''$ wherein $\rho_i' > \bar{\rho}_i$ and $\rho_i''<\bar{\rho}_i$, on account of $\int\rho_i dV - \int\bar{\rho}_i dV \equiv 0$, gives:

$$\int[\rho_i'(\mathbf{r},t) - \bar{\rho}_i(t)]dV' + \int[\rho_i''(\mathbf{r},t) - \bar{\rho}_i(t)]dV'' = 0. \qquad (1)$$

Hence it always follows the possibility to mark out *areas with processes of opposite directivity* in a heterogeneous system:

$$\int[d(\rho_i' - \bar{\rho}_i)/dt]dV' = -\int[d(\rho_i'' - \bar{\rho}_i)dt]dV'' = 0. \qquad (2)$$

The dissociation processes featuring $\bar{\rho}_i = 0$ are a particular case. This statement will appear to be a cornerstone of the modern thermodynamics. In its sence and generality it is in line with a well-known law of unity and struggle of opposites in materialistic dialectics. Its means so much for comprehending the specific character of heterogeneous processes that it should reasonably be endued with a status of special principle which we call the *principle of processes counterdirectivity: there are always subsystems with processes of opposite directivity existing within heterogeneous systems.* Hereafter it will be shown that these are such processes that cause the energy interconversion and finally the evolution of physical, biological and cosmological systems.

## 1.2. Exclusion of Hypotheses and Postulates from Theory Grounds

One of the most attractive features of the thermodynamic method has always been the possibility to obtain a great number of consequences of various phenomena as based on few primary principles ("the beginnings"), which are empirical laws in their character for the thermo-mechanical systems. Being consistently phenomenological (i.e. empirical), that method enabled to reveal general behavior of various processes without intrusion into their molecular mechanism and resort to simulation of structure and composition of a system under investigation. Therefore, it is not by pure accident that all the greatest physicists and many mathematicians of the last century (Lorenz, Poincaré, Planck, Nernst, Caratheodory, Sommerfeld, Einstein, Born, Fermi, Neiman, Landau, Zeldovich, Feynman, etc) in their investigations placed high emphasis on thermodynamics and, based on it, have obtained many significant results.

However, thermodynamics have presently lost its peculiar position among other scientific disciplines. It sounds now in increasing frequency that thermodynamics relates to real processes to the same degree as Euclidean geometry to the Egyptian land surveyors' work. Such a standpoint is not groundless. Classic thermodynamics is known to have always done with two primary postulates taken for its "beginnings" – the laws of excluded perpetual motion of the first and second kinds. Those principles have had the exclusion character and empirical status. However, classic thermodynamics restricted to those two laws appeared to have been unable to solve the problems that arose with its extension to phenomena of another nature. So in consideration of open systems exchanging substance with the environment, the entropy absolute value and the substance internal energy had to be known. To know the values, the third "beginning" would be needed as stating their becoming zero at the absolute zero of temperature. In-depth analysis of the thermodynamic logic structure in works of C. Caratheodory [22], T.A. Afanasjeva-Erenfest [23], A.A. Guhman [24] and their followers later led to the comprehension that the second law of thermodynamics would need to be split in two independent laws (existence and rise of entropy), as well as to realizing the important role of the equilibrium transitivity principle named the *zeroth law of thermodynamics* [7]. Starting to study non-equilibrium systems with irreversible processes running therein additionally required the L. Onsager's reciprocity principle sometimes named the *fourth law of thermodynamics* from the phenomenological positions. Further investigations have revealed the fundamental difference between statistical thermodynamics and phenomenological thermodynamics and the fundamental role that plays for the latter the equilibrium self-non-disturbance principle, which has been assigned a part of its "general beginning" [7]. Thus present day thermodynamics appears to be arisen from not two, but even seven beginnings! Meantime, the disputable consequences of thermodynamics are growing in number thus causing doubts in its impeccability as a theory. As R. Feynman wittily noted about this, "we have so many beautiful beginnings…but can't make ends meet nonetheless".

The law of excluded perpetual motion of the second kind being denied in open system thermodynamics [25], relativistic thermodynamics [8], spin system thermodynamics [26] excludes the possibility for modern thermodynamics to be based on the postulates of such a kind adopted for "the beginnings". The grave dissatisfaction investigators feel with such state of affairs has resulted in multiple attempts to build thermodynamics as based on other fundamental disciplines. This tendency has been most highlighted by A. Veinik [27] in his *thermodynamics of real processes* based on a number of postulates of quantum-mechanical character, by M. Tribus [28] in his *informational thermodynamics* based on the information theory formalism, and by C. Truesdall [29] in his *rational thermodynamics* topology-based. All these theories feature a denial of the consistently phenomenological (i.e. based on only empirical facts) approach to the theory of irreversible processes, which deprives them of the basic advantage intrinsic for the classic thermodynamic method – the indisputable validity of its consequences.

In our opinion, one of the reasons of such a situation is that thermodynamics has lost its phenomenological nature with considerations of statistical-mechanical character gaining influence in its conceptual basis. Whereas the founders of statistical mechanics strived to lay the thermodynamic laws into the foundation of statistical theories, a statement has become now common that phenomenological thermodynamics itself needs a statistical-mechanical substantiation (despite "there are much ambiguity" in the grounds of the statistical theories [30]). In particular, L. Onsager, the founder of the theory of irreversible processes (TIP), in order to substantiate the most fundamental concept of his theory – reciprocal relations, appealed to the principle of microscopic reversibility, the theory of fluctuations with a complementary postulate for linear character of their attenuation. All these statements evidently outspread beyond the thermodynamic applicability, therefore L. Onsager, not without reason, termed his theory *quasi-thermodynamics*.

Adoption of the *local equilibrium* hypothesis (I. Prigogine [16]) for a basis of TIP construction became even "further-reaching" assumption. This hypothesis assumes (a) equilibrium in the elements of heterogeneous systems (despite the absence of the necessary equilibrium criterion therein – termination of whatever macro-processes); (b) possibility to describe their status with the same set of parameters as for equilibrium (despite the actual use of additional variables – thermodynamic forces) and (c) applicability of the basic equation of thermodynamics to these elements (despite its inevitable transformation into inequality in case of irreversible processes). As a result, the existing theory of irreversible processes does not reach the rigor and completeness intrinsic for the classic thermodynamic method.

Striving for excluding postulates from the grounds of modern thermodynamics dictates the necessity to base modern thermodynamics on only those statements that are beyond any doubt and construed as axioms. These statements include, in particular, the *equilibrium self-non-disturbance axiom* reading that a thermodynamic system once having reached equilibrium cannot spontaneously leave it. Unlike the equilibrium self-non-disturbance principle (general law of thermodynamics), this axiom does not claim that a thermodynamic system, being isolated, reaches equilibrium for a finite time. The axiom just reflects the evident fact that processes in a system that has reached equilibrium may be generated by only impact applied to it from outside and are, therefore, never observed in isolated systems. Being a result of the experience accrued, this axiom excludes the possibility the macro-physical state of a system will vary as a result of short-term spontaneous deviations from equilibrium (fluctuations) caused by the micro-motion of the constituent particles. Indeed, if fluctuations do not cause any variation in the microscopic (statistical in their nature) parameters of the system, they can not be considered as an energy-involving process since the energy of the system remains invariable. Here lies the fundamental difference between modern thermodynamics and statistical physics – the latter does consider fluctuations as the object of investigation. At the same time the equilibrium self-non-disturbance axiom allows for existence of systems that omit the equilibrium state in their development since this axiom does not claim for relaxation time finiteness, which is hardly provable.

The *process distinguishability axiom* is another primary statement modern thermodynamics appeals to. It states there are processes existing and definable (by all experimental means) which cause system state variations as specific, qualitatively distinguishable and irreducible to any other ones. In classic thermodynamics these are isothermal, isobaric, adiabatic, etc processes. It will be shown hereinafter that these two axioms, in conjunction with experimental data underlying the energy conservation, are enough to construct a theory both internally and externally consistent and generalizing thermodynamics to transfer processes and conversion of energy in any forms.

## 1.3. Negation of Process and System Idealization outside the Framework of Uniqueness Conditions

Present-day thermodynamics has long outgrown the initial frames of the heat-engine theory and transmuted into a rather general macroscopic method for studying kinetics of various transfer processes in their inseparable connection with the thermal form of motion. However, it is still rooted in the conceptual system of equilibrium thermodynamics (thermostatics) distant from the transfer concept and in its body of mathematics going over into inequalities when considering real (non-static, irreversible) processes. Even in the current manuals on thermodynamics its construction quite often starts with describing the theory of ideal cycles and ideal gases as its working media. Such a "squared idealization" in the theory grounds themselves could not help creating problems in the further generalization of thermodynamics to systems differing from those idealized. First of all this applies to the scope of the correctives introduced at that into the primary notions of thermodynamics. Let us dwell on those absolutely necessary in view of changing to consideration of systems of a broader class. Such a correction relates, in the first turn, to the notion of *process* as itself because of existing in heterogeneous systems a specific class of *stationary irreversible processes* wherein local parameters of a system as the object under investigation remain invariable despite the flows of heat, substance, charge, etc available in this system. Striving to keep the primary notion of "process" as a *succession of state variations* makes it necessary to define this notion as *any space-time variation of macro-physical properties pertaining to an object of investigation*. Thereby the state variations associated with the spatial transfer of various energy forms are included in the notion of process.

Changing to consideration of real processes also demands to negate the process idealization as implied in such notions as the *quasi-static, reversible, equilibrium*, etc process. The notion of *process* as a sequence of state variations of an object under investigation and the notion of *equilibrium* as a state featuring the termination of whatever processes are mutually exclusive. To eliminate this contradiction is to recognize that any non-static (running with a finite rate) process means equilibrium disturbance and is, therefore, irreversible. The acknowledgment of the fact that any non-static (running with finite rate) process involves the equilibrium disturbance and thus is irreversible was a turning point in the logical structure of thermodynamics. That demanded, as will be shown hereinafter, to negate the first law of thermodynamics as based on the energy balance equation and to seek for other ways to substantiate the law of energy conservation. Being though somewhat previous, we can note that the solution to that problem was found by construing energy as the function of state for a spatially heterogeneous system and through its representation in terms of the parameters of that state without respect to the character of the processes in the system. As a result, all the remainder information about an object under investigation including the equation of its state and the kinetic equations of the processes running therein may be successfully attributed to the uniqueness conditions that thermodynamics imports "from outside" when applied to solving particular problems. In thermodynamics so constructed all the assumptions an investigator imposes on the uniqueness conditions (including the hypotheses on matter structure and process molecular mechanism, which simplify the preconditions for the equations of state and laws of transfer) do not affect the core of the theory itself, viz. those relations which follow from the mathematical properties of energy and other characteristic functions of system state.

Such a construction of thermodynamics is advisably to be started off with a notion of *action* introduced into thermodynamics long before the law of energy conservation was discovered. The action in mechanics is construed as something that causes the momentum variation $Md\mathbf{v}_o$, where $M$ – mass of the system, $\mathbf{v}_o$ – velocity of the mass center. According to the laws of mechanics the action value is expressed by the product of the force $\mathbf{F}$ and the duration of its action $dt$. This value is also called the *impulse of force*, N·s. A mechanical action is always associated with state variation, i.e. with process. Generalizing this notion to non-mechanical forms of motion the action will be construed as a *quantitative measure of a process associated with overcoming some* *forces*. The product of the action and the moving velocity **v** = *d***r**/*dt* of the object the force is applied to characterizes the amount of *work W*, J. The notion of work came to thermodynamics from mechanics (L. Carnot, 1783; J. Poncelet, 1826) where it was measured by the scalar product of the resultant force vector **F** and the induced displacement *d***r** of the object it was applied to (radius vector **r** of the force application center)

$$đW = \mathbf{F} \cdot d\mathbf{r} \qquad (3)$$

Thus work was considered as a quantitative measure of action from one body on another[1]. Later on forces were called mechanical, electrical, magnetic, chemical, nuclear, etc depending on their nature. We will denote the forces of the $i^{th}$ kind by $\mathbf{F}_i$ – according to the nature of this particular interaction form carrier. Forces are additive values, i.e. summable over the mass elements *dM*, bulk elements *dV*, surface elements *df*, etc. This means that in the simplest case they are proportional to some factor of their additivity $\Theta_i$ (mass *M*, volume *V*, surface *f*, etc) and accordingly called *mass, bulk, surface,* etc *forces*. Forces are also subdivided into *internal* and *external* depending on whether they act between parts (particles) of the system or between the system and surrounding bodies (the environment).

However, when considering non-equilibrium and, in particular, spatially heterogeneous media, another property of forces takes on special significance, viz. availability or absence of their resultant **F**. To clarify what conditions this availability or absence, it should be taken into consideration that from the positions of mechanics the work some force does is the only measure of action from one body (particle) on another. The forces of the $i^{th}$ kind generally act on the particles of different (the $k^{th}$) sort and hierarchical level of matter (nuclei, atoms, molecules, cells, their combinations, bodies, etc) possessing this form of interaction. Denoting the radius vectors of these elementary objects of force application by $\mathbf{r}_{ik}$ and the "elementary" force acting on them by $\mathbf{F}_{ik}$ gives that any $i^{th}$ action on a system as a whole is added of elementary works

$$đW_{ik} = \mathbf{F}_{ik} \cdot d\mathbf{r}_{ik} \qquad (4)$$

done on each of them ($đW_i = \Sigma_k \mathbf{F}_{ik} \cdot d\mathbf{r}_{ik} \neq 0$).

The result of such action will evidently be different depending on the direction of the elementary forces $\mathbf{F}_{ik}$ and the displacements $d\mathbf{r}_{ik}$ they cause. Let us first consider the case when the elementary forces $\mathbf{F}_{ik}$ cause the like-sign displacement $d\mathbf{r}_{ik}$ of the objects of force application (particles of the $k^{th}$ sort), i.e. change the position of the radius vector $\mathbf{R}_i$ for the entire set of the $k^{th}$ objects the elementary forces $\mathbf{F}_{ik}$ are applied to. In such a case $d\mathbf{R}_i = \Sigma_k d\mathbf{r}_{ik} \neq 0$ and the forces $\mathbf{F}_{ik}$ acquire the resultant $\mathbf{F}_i = \Sigma_k \mathbf{F}_{ik}$. This is the work done by mechanical systems and technical devices (machines) intended for the purposeful energy conversion from one kind into another. Therefore in technical thermodynamics such a work is usually called *useful external* or *technical* [31]. However, since in the general case such a work is done by not only technical devices, but biological, astrophysical, etc systems as well, we will call it just the *ordered* work and denote by $W^e$. The work of the $i^{th}$ kind is defined as the product of the resultant $\mathbf{F}_i$ and the displacement $d\mathbf{r}_i$ it causes on the object of its application:

$$đW_i^e = \mathbf{F}_i \cdot d\mathbf{r}_i . \qquad (5)$$

The ordered work process features its *vector character*. The work done at the uniform compression or expansion of a gas with no pressure gradients $\nabla p$ therein is another kind of work. Considering the local pressure *p* as a mechanical force acting on the vector element of the closed surface $df$ in the direction of the normal and applying the gradient theorem to the pressure forces resultant $\mathbf{F}_p$, gives:

$$\mathbf{F}_p = \int p df = \int \nabla p dV = 0. \qquad (6)$$

---

[1] Note that according to the dominating scientific paradigm only the interaction (mutual action) of material objects exists so that work is the *most universal* measure of their action on each other.

Thus the uniform compression work on an equilibrium (spatially homogeneous) system is not associated with the pressure forces resultant to be overcome, while the compression or expansion process itself is not associated with changing the position of the body as a whole. From the standpoint of mechanics where work has always been understood as an exclusively quantitative measure of energy conversion from one form into another (e.g., kinetic energy into potential one) this means that at the uniform compression the energy *conversion* process itself is absent. Due to the absence of the ordered motion of the $i^{th}$ object (its displacements $d\mathbf{r}_i = 0$) the work of such a kind will be hereinafter called *unordered* and denoted by $W^n$. This category should also include many other kinds of work not having a resultant, in particular, the work of uniformly introducing the $k^{th}$ substances (particles) or charge into the system, imparting relative motion momentum to the system components, etc. This category should further include heat exchange that is nothing but "micro-work" against chaotic intermolecular forces. As will be made certain hereinafter, the absolute value of the specific unordered forces $F_i/\Theta_i$ is construed as the *generalized potential* $\Psi_i$ (absolute temperature $T$, pressure $p$, electrical $\varphi$, chemical $\mu_k$ potential of the $k^{th}$ substances, etc). Thus the unordered work is done against forces not having a resultant. Therefore the unordered work process features the scalar character characterizing the transfer of energy in the same form (without energy conversion). This is the situation we encounter at the equilibrium heat or mass exchange and uniform cubic strain.

The work of dissipative character $W^d$ constitutes a special work category. This work is done by the ordered forces $\mathbf{F}_i$ against the so-called *dissipative forces* not having a resultant because of their chaotic directivity. Therefore the dissipative work features a mixed (scalar-vector) character, i.e. is associated with changing from *ordered* forms of energy to *unordered* ones.

Fig. 1.1 illustrated a work classification based on the force difference. energy conversion available in the ordered work processes is here indicated by superseding the subscript $i$ by the subscript $j = 1, 2, \ldots, n$ according to the nature of the forces being overcome. External work done (against environmental forces) is denoted by the superscript "*env*".

This work involves transferring a part of energy in a modified form to other bodies (environment). Internal work keeps the energy of the system invariable and involves its conversion from one form into another (as it occurs in oscillation processes or cyclic chemical reactions of Belousov-Zhabotinsky's type). Hereinafter this classification will underlie the classification of energy by its forms.

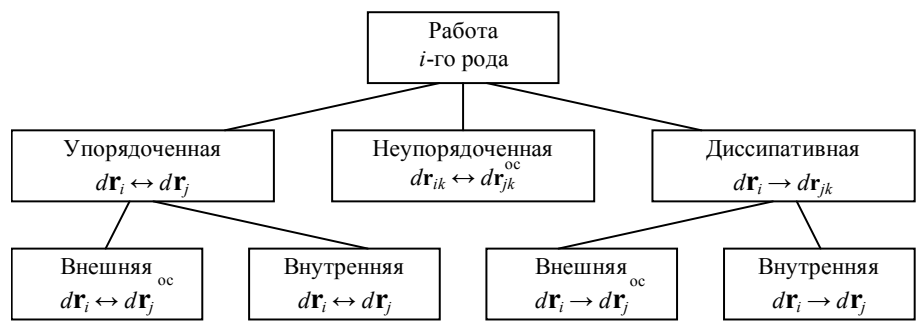

Fig. 1. Work Classification for Non-Equilibrium Systems

The term *heat* in the present-day technical literature is used in two meanings: as a *state* function (called briefly the *body heat*) and as a *process* function serving as a *quantitative* *measure of heat exchange* (and called briefly the *process heat*)[1]. This duality in construing heat appeared historically with considering heat as a chaotic form of motion (amongst such phenomena as light, sound, electricity, magnetism) and has remained notwithstanding multiple discussions. The conception of heat as a form of energy has been reflected in the notion of heat capacity of system. It has as well strengthened its position in the heat exchange theory (to the principle: a system can exchange only what it has). In non-equilibrium systems such an understanding of heat is dictated by a number of thermal effects caused by dissipation (friction, diathermic or induction heating, chemical transformations, etc). These heats are not supplied from outside either, though relate to *process*. However, in equilibrium systems of such a kind thermal effects are absent and heat becomes just a quantitative measure of heat exchange process. Therefore in equilibrium thermodynamics heat is often interpreted as the energy being transited from one body to another, i.e. something that is supplied from outside across the system borders, but not contained in the system itself.

Accepting the said duality for objectivity we will take into consideration both the body heat and the process heat denoting the former by $U_B$ (to avoid mishmash), while the latter – by $Q$ and applying the exact differential sign $d$ for infinitesimal increments of any state function (including $U_B$), while the inexact differential $đ$ – for the elementary heat amounts $đQ$ as process functions (C. Neuman, 1875). The same inexact differential sign will be applied for also the work $đW$ when it becomes dependent on the process path (i.e. becomes process function).

A specific kind of the energy exchange existing in the general case of open systems and associated with the substance (mass) exchange compels us to completely refuse the classic division of the energy exchange in such systems into *heat exchange* and *work*. The point is that putting substance into material medium always involves the so-called "input work" and the interchange of internal heat energy (body heat) between bodies. Therefore the notions of heat and work lose their sense "on the border where substance diffusion takes place" [28].

The impossibility to reduce the process heat to only "one of the forms of energy exchange" [32], as well as the existence of only one kind of energy exchange (mass exchange) in open systems forces absolutely rejecting the classic division of energy exchange into *heat* and *work* in non-equilibrium systems. The fundamental difference between the *ordered* and *unordered* work in non-equilibrium systems being perceived, heat exchange needs to be ascribed to the category of unordered work (against forces not having a resultant like the work of dissipative character). Interpreting the heat exchange as some *micro-work* against intermolecular forces directed in random way means the above notions are realized as different in their scale, while work is realized as a unitary quantitative measure of action from some material objects on other ones.

The abovementioned order of concepts clarifies the meaning and position of the notion of *energy*. The term *energy* (from Greek *activity*) was introduced into mechanics in the early XIX century by T. Young, an authoritative physicist, as a substitution for the notion of *living force* and meant the work which a system of bodies could do when decelerating or going over from a particular configuration into the "zeroth" one (adopted for the base). The energy was accordingly divided into kinetic $E^к$ и потенциальную $E^р$. The term *potential* meant that the energy could be realized in the form of work only with appearing the relative motion of interacting bodies, i.e. with changing their mutual position. The sum of kinetic and potential energies in an isolated (closed) system did not remain constant because of a known phenomenon of energy *dissipation* caused by unordered work done against the dissipation (friction) forces. Because of dissipation the real systems (with friction) spontaneously lost their capacity for external work. That meant the only thing, viz. the transition of energy as a microscopically ordered form of motion into the latent (microscopic) form of motion (interaction). Later, with thermodynamics appeared, that standpoint was supported by proving the internal energy $U$ as inherent to bodies. That allowed stating the law of conservation of total energy that was construed as the sum of kinetic $E^к$, potential $E^p$ and internal $U$ energies of an isolated system:

$$(E^к + E^p + U)_{isol} = \text{const.} \qquad (7)$$

---

[1] Thermodynamics. Terminology // under the editorship of I.I. Novikov. M.: AN USSR, 1973. Edition 85

However, in that case the notion of energy lost its primary sense of the capacity for external work as ensued from the word-group etymology of Greek εν (en) for *external* and εργον (ergon) for *action*. Indeed, according to the second law of thermodynamics the internal energy $U$ can not be entirely converted into work. For this reason the energy of real systems ceased to be determined by the amount of useful work done. And the work itself ceased to be the exact differential since became dependant on the process path and rate (dissipation conditioning), but not on exclusively the initial and final states of the system. To simplify the situation, mechanics was supplemented with a provisional notion of *conservative* system, where the sum of kinetic and potential energies could be considered as a value kept and dependant on exclusively the initial and final states of the system. However, in that case all the consequences of mechanics as having ensued from the energy conservation law were naturally restricted to only the conservative systems.

That engendered some ambiguity in the notion of energy, which has not yet been resolved. A reader is usually very surprised with not finding in physical guides and encyclopedia a definition of this notion more substantial than the philosophic category of *general quantitative measure of all kinds of matter*. As H. Poincaré bitterly noted, "we can say of energy nothing more but that something exists remaining invariable". Regarding the value that brings together all phenomena of the surrounding world such a definition is absolutely insufficient, the more so because not only energy remains invariable in isolated systems, but also mass, momentum, charge and angular momentum!

The definition modern thermodynamics offers for work through action and work interpreted as the only quantitative measure of action from some bodies on other ones allows returning to energy its simple and clear meaning as the capacity of a system to do work. However, now energy becomes a quantitative measure of all (ordered and unordered, external and internal, useful and dissipative) works a system can do. This approximates to the J. Maxwell's definition of energy as the "sum of all effects a system can have on the surrounding bodies". Next chapters will be dedicated to the substantiation of formal consistency and advisability of such an approach.

### 1.4. Compliance with the Adequacy Principle at System State Description

Changing over to non-equilibrium systems with spontaneous processes running therein needs to generalize the thermodynamic principle of *process classification* itself. The point is that the same state variations (e.g., heating of a body) in spatially heterogeneous systems may be caused by both the external heat exchange and appearing internal friction heat sources, chemical reactions, diathermic heating, magnetization reversal, etc. In the same way the cubic strain of a system can be induced by not only the compression work, but a spontaneous expansion into void as well. Hence processes in modern thermodynamics should be classified regardless of what causes whatever state variations – the external heat exchange or internal (including relaxation) processes.

In this respect modern thermodynamics differs from both physical kinetics that classifies processes by reasons causing them (distinguishing, in particular, concentration diffusion, thermal diffusion and pressure diffusion) and the heat exchange theory that distinguishes processes by the mechanism of energy transfer (conductive, convective and radiant). Processes in modern thermodynamics will be classified by their *consequences, i.e. by special state variations they cause as phenomenologically distinguishable and irreducible to others*. We will call such processes, for short, *independent*. These include, in particular, isochoric, isobaric, isothermal and adiabatic processes thermal physics considers. Here comes the *heat process* as well (K. Putilov, 1971), which we will construe as a variation of the body internal thermal energy $U_B$ regardless of what causes it – either heat exchange or internal heat sources. Other processes are also included, e.g., the system composition variation process that may be caused both by substance diffusion across the system borders and chemical reactions inside the system.

This statement is reasonable to be called for easy reference as the *distinguishability principle: there are independent processes existing – each causing system state variations as specific, phenomenologically distinguishable and irreducible to others.*

With the principle of process classification adopted as based on distinguishability of processes specific demands are made on choosing their "coordinate", i.e. a *physical value which variation is the necessary and sufficient criterion of running a particular process*. These demands consist in choosing only such a parameter as the process coordinate that *does not vary*, when other, also independent, processes are simultaneously running in the same space points. It is that approach wherefrom the requirement in classic thermodynamics follows for the invariability of entropy as the heat exchange coordinate in adiabatic processes as well as the requirement itself for the process reversibility, i.e. the absence of spontaneous entropy variations not connected with the external heat exchange.

The principle of classification of real processes by their consequences and the axiom of their distinguishability enable substantiating a quite evident though fundamental statement stipulating that *the number of independent coordinates conditioning the state of any (either equilibrium or non-equilibrium) modern thermodynamics system equals the number of degrees of its freedom, i.e. the number of independent processes running in the system.*

This statement (or theorem) is easily provable "by contradiction". Since a thermodynamic process is construed as variation of the properties of a system expressed in terms of state parameters, at least one of such parameters necessarily varies when processes are running. Let's assume that several state parameters necessarily vary when some independent process is running. Then these parameters will not evidently be independent, which violates the primary premise. Now let's assume that some coordinate of the system necessarily varies when several processes are running. Then these processes will not evidently be independent since they cause the same variations of the system properties – the fact that also violates the primary premise. We have nothing to do, but to conclude that *only one independent state coordinate corresponds to any* (equilibrium or non-equilibrium, quasi-static or non-static) *independent process*. Such coordinates are generally extensive variables since each of them defines, in the absence of other degrees of freedom, the energy of a system, which is an extensive value as well.

The proven statement defines the *necessary and sufficient* conditions for unique (deterministic) definition of state for whatever system. Therefore, it may be, for ease of reference, reasonably called the state *adequacy principle: the number of the independent parameters describing the state of some system is equal to the number of the independent processes running therein.* This principle makes it possible to avoid both the "under-determination" and "over-determination" of a system[1] as the main cause of the methodological errors and paradoxes of present-day thermodynamics [33]. The continuum state "under-determination" as resulting from the local equilibrium hypothesis adopted is, e.g., far from evidence. This hypothesis excludes the necessity of the gradients of temperature, pressure and other generalized potentials (i.e. thermodynamic forces) in the fundamental equation of non-equilibrium thermodynamics on the ground that the bulk elements are assumed to be equilibrium. On the other hand, the continuum "over-determination" due to the infinite number of degrees of freedom ascribed to it despite the finite number of macro-processes running therein is either not evident.

---

[1] I.e. the attempts to describe the system state by a deficient or excessive number of coordinates.

The theorem proven allows, in its turn, to concretize the notion of system *energodynamic state*, which is construed as a set of only such properties that are characterized by the set of state coordinates strictly defined in their number. This means that such system properties as color, taste, smell, etc, which are not characterized by state parameters quantitatively and qualitatively may not be considered as thermodynamic. This relates, in particular, to also the "rhinal", "haptic", etc number of freedom A. Veinik [27] arbitrarily introduced for a system.

One of the consequences of the determinacy principle consists in the necessity to introduce additional state coordinates for systems where, along with external heat exchange processes, internal (relaxation) processes are observed as tending to approximate the system to the equilibrium state. Without such variables introduced it is impossible to construct a theory covering the entire spectrum of real processes – from quasi-reversible up to critically irreversible.

### 1.5. Change to Absolute Reference System

The fact that relaxation vector processes (temperature, pressure, concentration, etc equalization) run in non-equilibrium systems requires introducing specific parameters of spatial heterogeneity characterizing the state of continuums as a whole. To do so, it is necessary, however, to find a way how to change over from the distribution functions of some extensive physical values $\Theta_i$ to the parameters of the system as a whole, which thermodynamics operates with. This change may be conducted in the same way as used in mechanics to change over from motion of separate points to system center-of-mass motion. To better understand such a change, let us consider an arbitrary continuum featuring non-uniform density distribution $\rho_i = \rho_i(\mathbf{r},t)$ of energy carriers[2] over the system volume $V$. Fig.2 illustrates the arbitrary density distribution $\rho_i(\mathbf{r},t)$ as a function of spatial coordinates (the radius vector of a field point $\mathbf{r}$) and time $t$. As may be seen from the figure, when the distribution $\Theta_i$ deviates from that uniform (horizontal line), some amount of this value (asterisked) migrates from one part of the system to other, which displaces the center of this value from the initial $\mathbf{r}_{i0}$ to a current position $\mathbf{r}_i$.

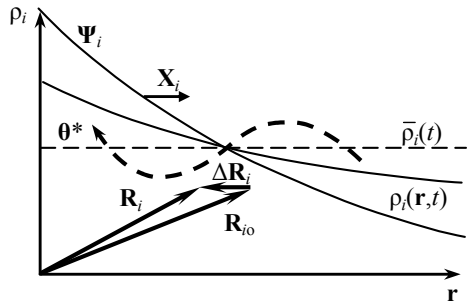

Fig.2. To Generation of Distribution Moment

Position of the center of a particular extensive value $\Theta_i$ defined by the radius vector $\mathbf{r}_i$ is given by a known expression:

$$\mathbf{r}_i = \Theta_i^{-1} \int \rho_i(\mathbf{r},t)\, \mathbf{r}\, dV, \quad (i = 1,2,\ldots,n) \qquad (8)$$

For the same system, but in a homogeneous state, the $\Theta_i$ center position $\mathbf{r}_{i0}$ may be derived if factoring $\rho_i = \bar{\rho}_i(t)$ in equation (2.1.1) outside the integral sign:

$$\mathbf{r}_{i0} = \int \bar{\rho}_i(t)\mathbf{r}\, dV = V^{-1} \int \mathbf{r}\, dV. \qquad (9)$$

Thus the state of a heterogeneous system features the emergence of specific "distribution moments" $\mathbf{Z}_i$ of the energy carriers $\Theta_i$:

$$\mathbf{Z}_i = \Theta_i(\mathbf{r}_i - \mathbf{r}_{i0}) = \int [\rho_i(\mathbf{r},t) - \bar{\rho}_i(t)]\, \mathbf{r}\, dV. \qquad (10)$$

---

[2] The Energy carrier is construed as a material carrier of the $i^{th}$ energy component, which quantitative measure is the physical value $\Theta_i$. So the mass $M_k$ of the $k^{th}$ substance is a carrier of the rest energy; the charge $\Theta_e$ – a carrier of the electrostatic energy of the system; the component momentum $M_k\mathbf{v}_k$ – a carrier of its kinetic energy, etc.

Expression (10) most evidently manifests that the parameters $\mathbf{Z}_i$ of spatial heterogeneity are additive values and summed up providing the $\bar{\rho}_i(t)$ value remains the same in various parts of a heterogeneous system. This follows from the conservation of integral (10) at its partition into parts with a volume $V' < V^{1)}$. However, these parameters become zero at "contraction" of the system to a material point, when $\rho_i(\mathbf{r},t) \to \bar{\rho}_i(t)$. This stands in absolute conformity with the degrees-of-freedom theorem because the processes of density redistribution $\rho_i(\mathbf{r},t)$ are absent in material points. And once again this confirms the fact that an entity of continuum elements considered as a system, non-equilibrium in whole, possesses additional degrees of freedom.

For any part of a homogeneous isolated system the $\mathbf{r}_{i0}$ value remains unvaried since running of any processes is herein impossible. Therefore the $\mathbf{r}_{i0}$ may be accepted for such systems as a reference point $\mathbf{r}$ or $\mathbf{r}_i$ and set equal zero ($\mathbf{r}_{i0} = 0$). In this case the vector $\mathbf{r}_i$ will define a displacement of the $\Theta_i$ center from its position for the system being in internal equilibrium state, and the moment of distribution of a particular value $\Theta_i$ in it will become:

$$\mathbf{Z}_i = \Theta_i \mathbf{r}_i \tag{11}$$

Herein the moment $\mathbf{Z}_i$ becomes an *absolute extensive measure of the system heterogeneity* with respect to one of the system properties – like such absolute parameters of classic thermodynamics as mass, volume, entropy, etc.

Let's show now, that all moments of distribution $\mathbf{Z}_i$ address in zero in homogeneous systems. With that end in view we will break non-uniform system $V$ on two parts with volume $V'$ and $V''$, in which limits the difference sign $\rho_i(\mathbf{r}, t) - \bar{\rho}_i(t)$ remains invariable. On fig. 1.2 them are areas, where $\rho_i(\mathbf{r}, t) > \bar{\rho}_i(t)$ and $\rho_i(\mathbf{r}, t) < \bar{\rho}_i(t)$. The moments of distribution $\mathbf{Z}_i'$, $\mathbf{Z}_i''$ and radiuses-vectors $\mathbf{r}_i'$, $\mathbf{r}_i''$ for these areas are defined by expression:

$$\mathbf{Z}_i = \int [\rho_i'(\mathbf{r},t) - \bar{\rho}_i(t)]\,\mathbf{r}\,dV' + \int [\rho_i''(\mathbf{r},t) - \bar{\rho}_i(t)]\,\mathbf{r}\,dV'' = \Theta_i'\mathbf{r}_i' + \Theta_i''\mathbf{r}_i'' \tag{12}$$

As $\Theta_i' = -\Theta_i''$, the moment of distribution $\mathbf{Z}_i$ can give a kind of dipole moment:

$$\mathbf{Z}_i = \Theta_i'(\mathbf{r}_i' - \mathbf{r}_i'') = \Theta_i^* \Delta \mathbf{r}_i, \tag{13}$$

where $\mathbf{Z}_i$, $\Theta_i^* = \Theta_i' = -\Theta_i''$; $\Delta \mathbf{r}_i$ – the sizes similar on sense of doublet moment, dipole charge and a dipole shoulder. From here follows, that the distribution moments, similarly to the dipoles moments, address in zero with disappearance of spatial heterogeneity of system ($\Delta \mathbf{r}_i = 0$) irrespective of a parameter reference mark $\Theta_i^*$.

In case of discrete systems the integration over system volume will be replaced by the summation with respect to elements $d\Theta_i$ of the $\Theta_i$ value:

$$\mathbf{Z}_i = \Theta_i \mathbf{r}_i = \Sigma_i \mathbf{r}_i\,d\Theta_i, \tag{14}$$

where $\mathbf{r}_i$ – radius vector of the element $d\Theta_i$ center. Therefore expressions (14) through (13) remain valid for also the systems with discrete distribution of charges, poles, elementary particles, etc. Only the geometrical meaning of the $\Delta \mathbf{r}_i$ vector changes; for symmetrical distributions the vector is defined by the sum of the displacements $\Delta \mathbf{r}_i$ of all elements $d\Theta_i$. This may be instantiated by the centrifugal "shrinkage" of the particles' momentum flow in moving liquid when forming turbulent or laminar fluid-velocity profiles in channels ("boundary layer" formation and build-up).

---

[1)] With symmetrical density $\rho_i(\mathbf{r},t)$ distributions for whatever parameter, e.g., fluid-velocity profiles in tubes, expression (1.5.3) should be integrated with respect to annular, spherical, etc layers with $V' > 0$, wherein the function $\rho_i(\mathbf{r},t)$ is monotone increasing or decreasing.

Explicitly taking into account the spatial heterogeneity of systems under investigation is decisive in further generalization of the thermodynamic investigation method to non-equilibrium systems. As a matter of fact, this is the spatial heterogeneity (heterogeneity of properties) of natural objects that causes various processes running in them. This implies the exclusive role the distribution moments $\mathbf{Z}_i$ play as a measure for deviation of a system in whole from internal equilibrium of the $i^{th}$ kind. Introducing such parameters allows precluding the major drawback of non-equilibrium thermodynamics, viz. lack of extensive variables relating to the gradients of temperature, pressure, etc. Classic thermodynamics is known to have crystallized into an independent discipline after R. Clausius succeeded in finding a coordinate (entropy) related to temperature in the same way as pressure to volume and thus determinately described the simplest thermo-mechanical systems. The distribution moments $\mathbf{Z}_i$ play the same part in modern thermodynamics coming into being. As will be shown later, these relate to the main parameters introduced by non-equilibrium thermodynamics – thermodynamic forces, in the same way as the generalized potentials to the coordinates in equilibrium thermodynamics. These are the distribution moments which make the description of heterogeneous media a deterministic one thus enabling introducing in natural way the concept of generalized velocity of some process (flow) as their time derivatives. They visualize such parameters as the electrical displacement vectors in electrodynamics and generalize them to phenomena of other physical nature. In mechanics the $\mathbf{Z}_i$ parameters have the dimension of action ($\Theta_i$ – momentum of a body, $\mathbf{r}_i$ – its displacement from equilibrium position), imparting physical meaning to this notion. These are the parameters which allow giving the analytic expression to the system working capacity having thus defined the notion of system energy. Using such parameters provides a clear view of the degree of system energy order, enables proposing a universal criterion of the non-equilibrium system evolution, etc. Paraphrasing a M. Planck's statement regarding entropy one may positively say that the distribution moments are exactly the parameters entire non-equilibrium thermodynamics is "standing and falling" with.

### 1.6. Counterdirectivity of Processes Allowed for When Finding Their Coordinates

As follows from expressions (9) and (10), the distribution moment $\mathbf{Z}_i$ contain vectors of displacement $\mathbf{r}_i$, each of which can be expressed product of a basic (individual) vector $\mathbf{e}_i$, characterising its direction, on module $R_i = |\mathbf{R}_i|$ this vector. Therefore the complete variation of the displacement vector $\mathbf{R}_i$ may be expressed as the sum of two summands:

$$d\mathbf{r}_i = \mathbf{e}_i dr_i + r_i d\mathbf{e}_i = \mathbf{e}_i dr_i + r_i d\mathbf{\varphi}_i \times \mathbf{e}_i, \qquad (15)$$

where the augend $\mathbf{e}_i dr_i$ characterizes elongation of the vector $\mathbf{r}_i$, while the addend $r_i d\mathbf{e}_i$ – its turn.

Let us express now the $d\mathbf{e}_i$ value characterizing the variation of the distribution moment direction in terms of an angular displacement vector $\mathbf{\varphi}$ normal to the plane of rotation formed by the vectors $\mathbf{e}_i$ and $d\mathbf{e}_i$. Then the $d\mathbf{e}_i$ will be defined by the external product $d\mathbf{\varphi}_i \times \mathbf{e}_i$ of vectors $d\mathbf{\varphi}_i$ and $\mathbf{e}_i$, so the addend in (15) will be $\Theta_i r_i d\mathbf{e}_i = d\mathbf{\varphi}_i \times \mathbf{Z}_i$. Hence, expression of full differential of the distribution moments looks like:

$$d\mathbf{Z}_i = d(\Theta_i \mathbf{e}_i r_i) = \mathbf{r}_i d\Theta_i + \Theta_i \mathbf{e}_i dr_i + d\mathbf{\varphi}_i \times \mathbf{Z}_i, \qquad (16)$$

According to the degrees-of-freedom theorem this means that any state function describing a heterogeneous system in whole are generally defined by also the full set of variables $\Theta_i$, $r_i$ and $\mathbf{\varphi}_i$. Since further resolution of the vector $\mathbf{Z}_i$ is impossible, expression (16) indicates there are three categories of processes running in heterogeneous media, each having its own group of independent variables. The first-category processes running at $\mathbf{r}_i$ = const involve the uniform variation of the physical value $\Theta_i$ in all parts of the system. Such processes resemble the uniform rainfall onto an irregular (in the general case) surface. Here comes, in particular, the pressure field altered in liquid column with variation of free-surface pressure. These processes also cover phase transitions in emulsions, homogeneous chemical reactions, nuclear transformations and the similar scalar processes providing the composition variations they induce are the same in all parts of the system. We will call them hereinafter the *uniform processes* regardless of what causes the increase or decrease in amount of whatever energy carrier $\Theta_i$ (and the momentum associated) – either the external energy exchange or internal relaxation phenomena. These processes comprise, as a particular case, the reversible (equilibrium) processes of heat exchange, mass exchange, cubic strain, etc, which, due to their quasi-static nature, practically do not disturb the system spatial homogeneity. The counterdirectivity of these processes shows up in the opposite variation of the parameters $\Theta_i$ in the system and the environment.

Processes described by the addend in (12) run with the $\Theta_i$ parameters being constant and consist in their redistribution among the parts (zones) of a heterogeneous system. They involve decreasing, e.g., the entropy $S'$, mass $M'$, its momentum $\mathbf{P}'$, its volume $V'$, etc, in some parts of the system and by increasing the same in other parts. Such processes are associated with the $\Theta_i$ value center position variation $\mathbf{r}_i$ within the system and resemble the migration of fluids from one part of a vessel into another. Therefore we will call them the *redistribution* processes. The electrical displacement vector $\mathbf{D} = \Theta_e \Delta \mathbf{r}_e$ is one of such moments with $\Theta_e$ as electrical charge and $\Delta \mathbf{r}_e$ as displacement of its center.

Such processes are always non-equilibrium even if they run infinitely slowly (quasi-statically) since the system remains spatially heterogeneous in this case. State modifications of such a kind are caused by, e.g., the useful external work of external forces, the non-equilibrium energy exchange processes that induce non-uniform variation of the $\Theta_i$ coordinates inside the system, and the vector relaxation processes involving equalization of temperature, pressure, chemical and other system potentials. All processes of such a kind feature a directional (ordered) character, which distinguishes the useful work from the work of uniform (quasi-static) introduction of substance, charge, etc, or the expansion work. Their counterdirectivity shows up in the opposite displacement of centers for the values $\Theta_i'$ and $\Theta_i''$ in the subsystems $V'$ and $V''$. According to (1.3.2) the coordinates of the processes pertaining to this category are understood as the displacement vectors $\mathbf{r}_i$. These coordinates should be attributed to the *external parameters* of the system since they characterize the *position* of the energy carrier $\Theta_i$ center in whole relative to external bodies (the environment) just as the center of mass $\mathbf{r}_m$ of the system or its center of inertia $\mathbf{r}_w$.

There are also the processes of *reorientation* of magnetic domains, electrical and magnetic dipoles, axes of rotation of bodies, etc., running in a number of systems, e.g., in ferromagnetic materials. The micro-world manifests them in, e.g., the unified spin-orientation arrangement' the macro-systems – in the spontaneous magnetization of ferromagnetic materials, while the mega-world – in the close-to-equatorial plane alignment of the galaxies' spirals, asteroidal belts, orbits of the primary planets and their satellites, etc. The counterdirectivity of these processes shows up in the opposite displacement of the vector $\mathbf{Z}_i$ termini at vector turn. The systems with processes of such a kind will hereafter be called, for short, *oriented*. These include also the bodies with shape anisotropy. The reorientation processes are not reducible to the transfer and redistribution processes either. This means that the coordinate of such kind a process is a variation of the angle $\varphi_i$ characterizing the orientation of distribution moment $\mathbf{Z}_i$ of the system as a whole.

Thus, all processes running in heterogeneous systems may be broken down into three groups: *uniform, redistribution* and *reorientation processes*, which coordinates are, respectively, variables $\Theta_i$, $r_i$ and $\varphi_i$. This fundamentally distinguishes modern thermodynamics from classic thermodynamics and the theory of irreversible processes, where the state of a system is defined by exclusively a set of thermostatic variables $\Theta_i$.

For lack of processes of reorientation $\mathbf{Z}_i = \mathbf{Z}_i(\Theta_i, \mathbf{r}_i)$ and expression (16) becomes:

$$d\mathbf{Z}_i = \mathbf{r}_i\, d\Theta_i + \Theta_i d\mathbf{r}_i, \quad (17)$$

resulting in:

$$\Theta_i = \nabla \cdot \mathbf{Z}_i \text{ and } \rho_i = \nabla \cdot \mathbf{Z}_{iV}, \quad (18)$$

where $\mathbf{Z}_{iV} = \partial \mathbf{Z}_i/\partial V$ – distribution moment in the system unit volume.

The undertaken expansion of the space of variables by introducing the vectors of displacement $\mathbf{r}_i$ makes it possible to cover not only quantitative, but as well *qualitative* variations of energy in various forms. The fact that *vector processes* run in systems along with *scalar processes* means that both the *ordered* $W^e$ and *unordered* $W^a$ works are generally done in such systems. It becomes clear that the irreversibility of real processes associated with the energy dissipation (i.e. with losing the capacity for ordered work) becomes apparent in the process «*scalarization*», i.e. in losing vector character of the process. Furthermore, a possibility appears to further distinguish between the *energy transfer* processes (i.e. the energy transfer between bodies in the same form) and the *energy transformation* processes (i.e. the energy conversion from one form into another)[1].

This is enough in principle to construct a unitary theory of real processes enabling investigation of any systems (simplex and complex, closed and open, homogeneous and heterogeneous, isolated and non-isolated, tending to and omitting equilibrium) not outstepping the strict applicability of its primary concepts [33].

---

[1] As will be shown hereinafter, the Energy transfer is associated with unordered work done, whereas the Energy conversion – with ordered work.